\documentclass[sigconf]{acmart}
\usepackage{multirow}
\usepackage{subfigure}
\usepackage{bm}
\usepackage{graphicx}
\usepackage{enumitem}
\AtBeginDocument{%
  }

\copyrightyear{2023} 
\acmYear{2023} 
\setcopyright{acmlicensed}
\acmConference[MM '23]{Proceedings of the 31st ACM International Conference on Multimedia}{October 29-November 3, 2023}{Ottawa, ON, Canada}
\acmBooktitle{Proceedings of the 31st ACM International Conference on Multimedia (MM '23), October 29-November 3, 2023, Ottawa, ON, Canada}
\acmPrice{15.00}
\acmDOI{10.1145/3581783.3611915}
\acmISBN{979-8-4007-0108-5/23/10}

\acmSubmissionID{972}

\begin{document}


\title{Mamba: Bringing Multi-Dimensional ABR to WebRTC}

\author{Yueheng Li}
\authornote{Both authors contributed equally to this research.}
\affiliation{%
  \institution{Nanjing University}
  \city{Nanjing}
  \country{China}}
\email{yueheng.li@smail.nju.edu.cn}

\author{Zicheng Zhang}
\authornotemark[1]
\affiliation{%
  \institution{Nanjing University}
  \city{Nanjing}
  \country{China}}
\email{zichengzhang@smail.nju.edu.cn}

\author{Hao Chen}
\authornote{Corresponding author: Hao Chen.}
\affiliation{%
  \institution{Nanjing University}
  \city{Nanjing}
  \country{China}}
\email{chenhao1210@nju.edu.cn}

\author{Zhan Ma}
\affiliation{%
  \institution{Nanjing University}
  \city{Nanjing}
  \country{China}}
\email{mazhan@nju.edu.cn}


\renewcommand{\shortauthors}{Y. Li et al.}
%
\begin{abstract}
Contemporary real-time video communication systems, such as WebRTC, use an adaptive bitrate (ABR) algorithm to assure high-quality and low-delay services, e.g., promptly adjusting video bitrate according to the instantaneous network bandwidth. However, target bitrate decisions in the network and bitrate control in the codec are typically incoordinated and simply ignoring the effect of inappropriate resolution and frame rate settings also leads to compromised results in bitrate control, thus devastatingly deteriorating the quality of experience (QoE). To tackle these challenges, Mamba, an end-to-end multi-dimensional ABR algorithm is proposed, which utilizes multi-agent reinforcement learning (MARL) to maximize the user's QoE by adaptively and collaboratively adjusting encoding factors including the quantization parameters (QP), resolution, and frame rate based on observed states such as network conditions and video complexity information in a video conferencing system. We also introduce curriculum learning to improve the training efficiency of MARL. Both the in-lab and real-world evaluation results demonstrate the remarkable efficacy of Mamba.
\end{abstract}




\maketitle

\section{Introduction}
\label{sec:intro}
The exponential surge of real-time video communication (RTVC) applications, such as video conferencing, remote sharing, and online education, has been observed in recent years. With the emergence of COVID-19, these applications have even become a part of our daily lives, resulting in notable consumption of internet video traffic (e.g., exceeding 17\%~\cite{video_traffic2022}), and significant revenue returns (e.g., billions of dollars worldwide~\cite{video_market2022}).


In video-on-demand (VoD) applications, content providers commonly encode the videos at different bitrate levels in advance and resort to adaptive bitrate (ABR) technology to combat the underlying network fluctuations for quality optimization~\cite{Pensieve,MPC}. However, in the RTVC applications, bitrate adaption becomes more challenging due to the unavailable actual compressed video bitrate.
The ABR strategy implemented in WebRTC~\cite{webrtc}, a typical RTVC framework, encompasses two primary processes: congestion control (or bandwidth estimation/target bitrate decision\footnote{In this paper, we use these three terms interchangeably to refer to the same algorithm.}) and video bitrate control. The former involves estimating the available bandwidth by analyzing network state metrics such as packet loss rate and round-trip time (RTT) interval. The estimated bandwidth then serves as the target bitrate for the application-layer encoder to encode a video. To make the video bitrate match the target bitrate, the encoder adaptively determines the quantization parameter (QP) for each frame using the built-in rate control algorithms~\cite{h264}.
Meanwhile, WebRTC dynamically adjusts the resolution or frame rate to compensate for the inaccurate bitrate control of the encoder.
However, such a paradigm suffers from several serious problems, making the performance of existing RTVC applications far from optimal. Firstly, the incoordination of the transport layer and application layer weakens the effectiveness of the ABR strategy (see Figure~\ref{fig:WebRTC_compare}). This is brought about by the limited information sharing and mismatch of decision granularity between the congestion control and bitrate control~\cite{salsify, concerto, palette}. Moreover, existing frame rate or resolution adaptation strategies use preset rules in the form of fixed heuristics, which cannot adapt to diverse video content (see Figure~\ref{fig:switch}). Finally, the unsophisticated combination of the rate control algorithm in the encoder and external adaptation policy for frame rate and resolution fails to achieve a perfect collaboration, consequently degrading the final performance (see Figure~\ref{fig:WebRTC_vs_Mamba}). Recently, reinforcement learning (RL)-based congestion control methods~\cite{concerto, ars, loki} have emerged with the aim to improve the accuracy of bandwidth estimation at the transport layer. Nevertheless, they do not step out of the aforementioned paradigm and thus share the same drawbacks.

This paper proposes a novel end-to-end multi-dimensional adaptive bitrate (MABR) framework for RTVC applications, named Mamba, which considers both network conditions and video content complexity to decide the encoding factors (e.g., QP, frame rate, and resolution) that jointly determine the video bitrate. The proposed Mamba avoids the use of explicit target bitrate, which may lead to incoordination between the transport and application layers. Instead, it directly optimizes the user's QoE in an end-to-end manner based on network conditions and video content complexity information. To this aim, we model the joint adaption of the encoding factors as a multi-agent decision problem and devise a multi-agent reinforcement learning (MARL) approach to learn the appropriate combination of the encoding factors through exploring the environment. During the training, curriculum learning is introduced to improve the training efficiency and asymptotic performance (model performance after reaching convergence) of Mamba. 

In summary, our contributions are as follows:
\begin{itemize}[leftmargin=*]
    \item Based on deep analysis of the ABR strategy used by WebRTC (§\ref{ssec:abr_in_webrtc}), we identify the limitations of widely-used WebRTC on providing a satisfactory QoE (§\ref{ssec:observ}) and further propose an end-to-end multi-dimensional ABR framework to overcome these limitations (§\ref{sec:ma_design}).
    \item To the best of our knowledge, this is the first work that employs multi-agent reinforcement learning to adaptively adjust the joint encoding factors including QP, frame rate, and resolution, to optimize the QoE of RTVC applications based on observed network conditions and video content complexity (§\ref{sec:design}).
    \item We have integrated Mamba into the WebRTC framework and evaluated its performance by comparing it to state-of-the-art ABR algorithms. Extensive experimental results in real-world field tests indicate that Mamba outperforms the native WebRTC with improvements in video quality of 16.9\% while maintaining a similar frame rate and delay (§\ref{sec:evaluation}).
\end{itemize}

\section{Background and Motivation}
\label{sec:motivation}

Nowadays, RTVC has emerged as the principal medium for individuals to engage in social, recreational, and professional activities. As the first open-source and consistently updated RTVC framework, WebRTC~\cite{webrtc} has gained significant traction and has been supported natively by various popular browsers such as Chrome, Firefox, and Safari. The technologies employed by WebRTC, e.g., congestion control, adaptive bitrate, and packet loss recovery, have been adopted in various other RTVC applications, including low-latency live streaming, cloud gaming, and video conferencing~\cite{ars, 10.1145/3199524.3199534}. This work presents a novel ABR framework for the optimization of existing WebRTC systems. Nonetheless, the proposed ABR framework holds potential for broader applicability across all RTVC applications.

\subsection{Adaptive Bitrate in WebRTC}
\label{ssec:abr_in_webrtc}
The ABR policy in WebRTC can be broadly divided into two primary cascaded processes: target bitrate decision and bitrate control. In the following, we provide a concise summary of these two parts.

\textbf{Target bitrate decision.}
WebRTC adopts the Google Congestion Control (GCC)~\cite{gcc} algorithm to estimate current network capacity and further employs the estimated bandwidth as the target bitrate for video encoding. The GCC algorithm comprises two controllers: the packet loss-based controller and the delay-based controller. The former undertakes three different actions on the sending bitrate based on the packet loss rates found in RTCP packets: decrease, maintain, or increase. Meanwhile, the latter first filters round trip time (RTT) variations to divide the network into three states: overuse, underuse, and normal, and then performs corresponding decrease, increase, or maintenance operations on the sending bitrate accordingly. The final sending bitrate is determined by the minimum value of the two controllers' outcomes.

\begin{figure}[t]
\setlength{\abovecaptionskip}{-0.1cm}
    \centering
    \subfigure[Resolution adjustment off]
    {
        \label{sfig:webrtc_wo_res}
        \begin{minipage}[ht]{.48\linewidth}
        \centering
            \includegraphics[width=\linewidth]{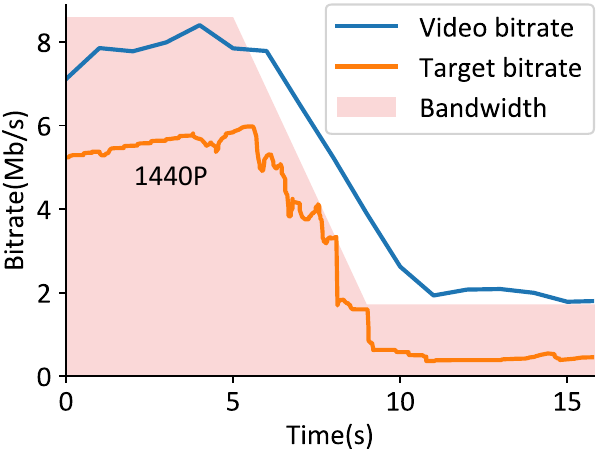}
        \end{minipage}
    }\hspace{-5pt}
    \subfigure[Resolution adjustment on]
    {
        \label{sfig:webrtc_w_res}
        \begin{minipage}[ht]{.48\linewidth}
        \centering
            \includegraphics[width=\linewidth]{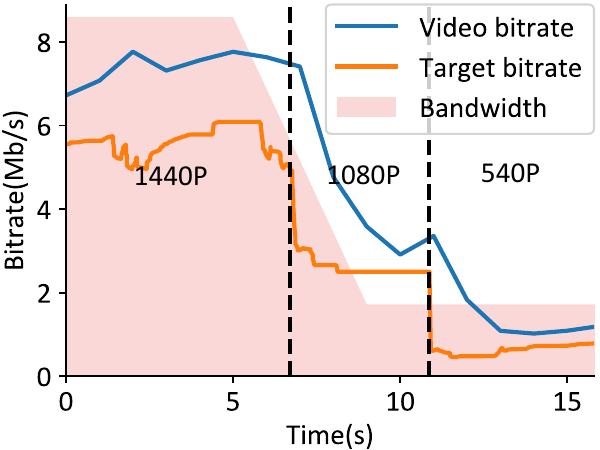}
        \end{minipage}
    }
    \caption{The mismatch between target bitrate and video bitrate using WebRTC in both resolution adjustment on and off modes, reflects the incoordination issue between transport and application layers.}
    \label{fig:WebRTC_compare}
    \vspace{-12pt}
\end{figure}

\textbf{Bitrate control.}
Initially, WebRTC uses the estimated network bandwidth generated by GCC as the target encoding bitrate, at which the video encoder is called for video compression using its built-in rate control algorithm (e.g., in variable bitrate (VBR) or constant bitrate (CBR) mode). However, it is realized by adjusting the QP of each frame only, without considering the resolution and frame rate, which also determine the final video bitrate as reported in~\cite{ma2011modeling}.   
Therefore, WebRTC also integrates a resolution adjustment module and a frame rate adjustment module to further enhance its bitrate control. The resolution adjustment module builds a fixed mapping table with separate entries for all bitrate ranges and corresponding resolution configurations. By looking up this table, the resolution is determined for a given target encoding bitrate. On the other hand, the frame rate adjustment module employs a leaky bucket algorithm to skip the next frame(s) whenever the actual video bitrate surpasses the target bitrate. It is worth noting that for the current version of WebRTC, the resolution adjustment module and the frame rate adjustment module cannot be activated concurrently, and the former is enabled by default. In this paper, we keep the default settings in the implementation of WebRTC. 

\subsection{Observations}
\label{ssec:observ}

In this section, we set up preliminary experiments to evaluate the ABR strategy of WebRTC and yield a set of remarkable findings. These experiments were conducted on our testbed which is detailed in §\ref{ssec:eva_methodology}. 

\subsubsection{Incoordination between transport and application layers}
\label{sssec:incordination}
As introduced in §\ref{ssec:abr_in_webrtc}, WebRTC's ABR policy is composed of two separate algorithms (i.e., target bitrate decision and bitrate control) which are deployed at the transport layer and the application layer respectively, and short of shared information except for the target encoding bitrate. This results in coordination issues, including adaptation lag and optimization objective misalignment, as observed in previous works~\cite{concerto, palette}. To demonstrate these issues, we conducted an experiment by simulating network congestion events with the setting of varying available bandwidths and streaming a video that is randomly chosen from the YouTube User Generated Content (YT-UGC~\cite{yt-ugc}) dataset following the WebRTC protocol.

The performances of WebRTC with resolution adjustment on and off, are illustrated in Figure~\ref{fig:WebRTC_compare}. Figure~\ref{sfig:webrtc_wo_res} reveals a significant discrepancy between the target encoding bitrate and the actual video bitrate for WebRTC, primarily due to the inaccurate bitrate control by the encoder in low-delay encoding mode. The target bitrate generated by GCC decreases rapidly at around 7s, while the video bitrate generated by the encoder fails to follow it closely. This bitrate offset further impairs the accuracy of bandwidth estimation for GCC (after 10s). As a result, prolonged overshooting occurs (after 10s) due to ineffective bitrate control of WebRTC, which seriously deteriorates the user's QoE. Although WebRTC in later versions has introduced a dynamic resolution adaptation scheme to mitigate the incoordination problem, it is still far from satisfactory. As shown in Figure~\ref{sfig:webrtc_w_res}, the mismatch between target bitrate and video bitrate still exists, and overshooting is observed from 7s to 13s. We desire an end-to-end adaptation policy that straightforwardly performs bitrate control based on network conditions and video information, bypassing inaccurate bandwidth estimation.

\begin{figure}[t]
    \centering
    \setlength{\abovecaptionskip}{-0.cm}
    \includegraphics[width=0.95\linewidth]{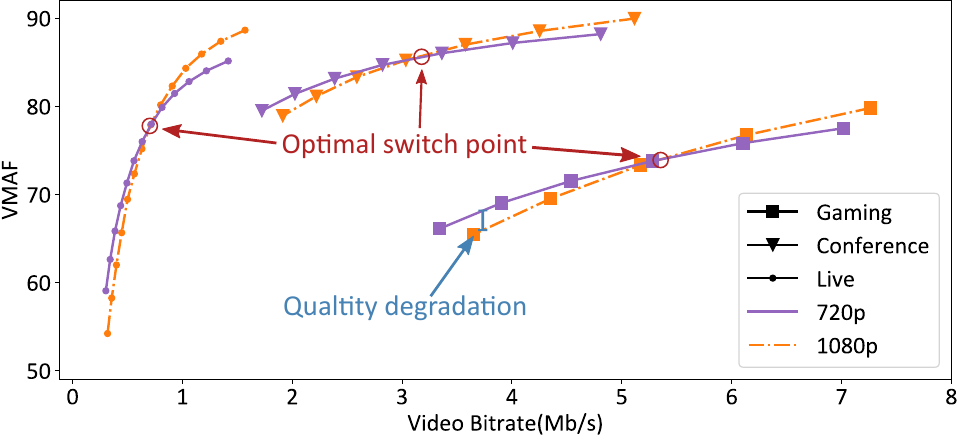}
    \caption{The optimal resolution switch points (red circles) vary for different types of videos. A sub-optimal switch could lead to a noticeable quality degradation (blue mark).}
    \label{fig:switch}
    \vspace{-12pt}
\end{figure}

\subsubsection{Limitation of fixed rules-based resolution adaptation}
Even though WebRTC introduces a resolution adaptation policy to enhance the efficiency of bitrate control, it is fixed rules-based, which does not generalize to different video contents. We conducted an experiment to demonstrate this issue by selecting three video contents with different complexity from the YT-UGC dataset: live streaming, video conferencing, and games. These videos were encoded with 1080P and 720P resolutions at different bitrates. VMAF~\cite{VMAF} is employed to evaluate picture quality. As depicted in Figure~\ref{fig:switch}, the optimal resolution switching point for different content varies considerably, and a sub-optimal switching point leads to a drop in VMAF by up to 5. This emphasizes the importance and necessity of a more effective resolution adaption strategy.

\begin{figure}[t]
    \centering
    \setlength{\abovecaptionskip}{-0.cm}
    \includegraphics[width=0.95\linewidth]{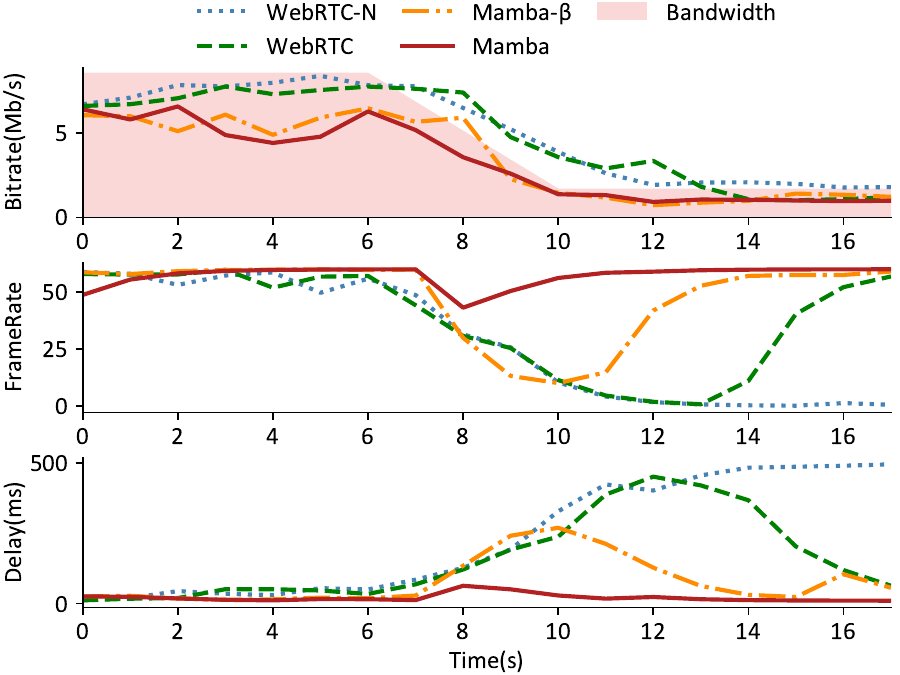}
    \caption{Performance comparison using WebRTC-N, default WebRTC, Mamba-\bm{$\upbeta$}, and full Mamba. The outstanding performance of Mamba confirms the importance of a joint adaption of QP, resolution, and frame rate.}
    \label{fig:WebRTC_vs_Mamba}
    \vspace{-12pt}
\end{figure}

\subsubsection{Importance of joint adaption of QP, resolution, and frame rate}
In addition to the first preliminary experiments, we conducted another experiment to show the impact of the above-mentioned WebRTC's limitations on QoE metrics, including video bitrate, playback frame rate, and delay. Using the same environment configuration as §\ref{sssec:incordination}, we compare the results using WebRTC with resolution adaptation off (WebRTC-N), default WebRTC, Mamba with the joint adaption of QP and resolution (Mamba-$\upbeta$), and full Mamba (with a joint adaptation of QP, resolution, and frame rate). As shown in Figure~\ref{fig:WebRTC_vs_Mamba}, the video for WebRTC-N almost entirely freezes after 12 seconds, while that for WebRTC eventually resumes normal playback. This again highlights the significance of dynamic resolution adaptation. Mamba-$\upbeta$ achieves a faster recovery time and lower delay than WebRTC, which verifies the effectiveness of the end-to-end adaptation policy. Furthermore, Mamba applies an additional frame rate adaptation to reduce the video bitrate more quickly when facing network congestion, avoiding excessive packet losses. By actively downgrading the encoding frame rate, Mamba increases the likelihood that the received video frames can be decoded and achieves a higher playback frame rate and lower delay, resulting in a higher QoE. This finding reveals that a joint adaptation of QP, resolution, and frame rate can further improve ABR performance.

\section{Multi-Dimensional ABR Algorithm}
\label{sec:ma_design}
\begin{figure}[t]
    \centering
    \setlength{\abovecaptionskip}{-0.cm}
    \includegraphics[width=\linewidth]{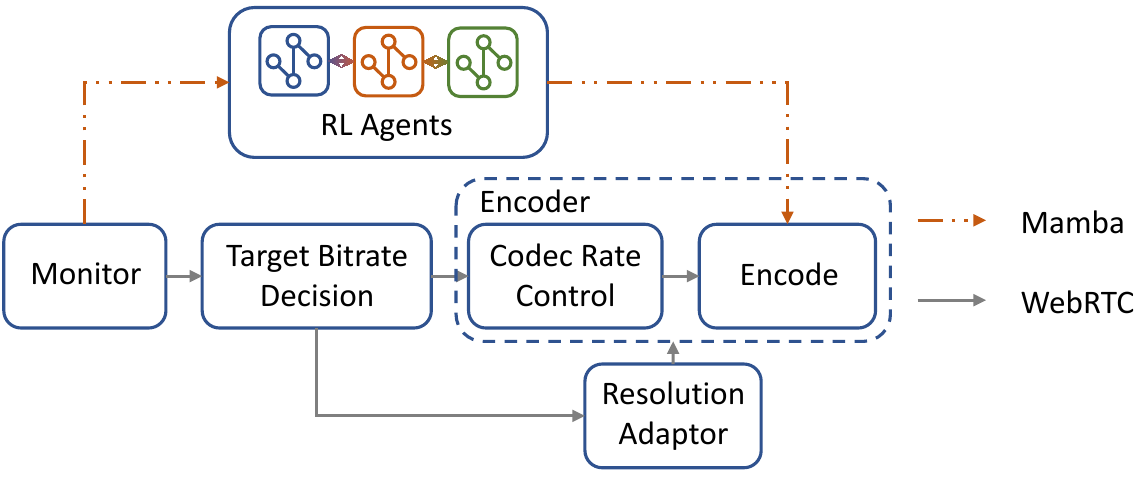}
    \caption{The key brilliance of Mamba compared to default WebRTC.}
    \label{fig:compare}
    \vspace{-16pt}
\end{figure}

Based on the experimental analysis presented in §\ref{ssec:observ}, we learn that the existing WebRTC framework cannot ensure a satisfactory QoE due to incoordination between the transport and application layers and the unsophisticated combination of multi-dimensional bitrate control policies. In this paper, we propose Mamba, an end-to-end multi-dimensional ABR framework, designed to learn end-to-end adaptation policies that directly map the observations (e.g., network conditions, and video information) to the best configurations of encoding factors including QP, resolution, and frame rate for QoE optimization. A comparison of the ABR framework between Mamba and WebRTC is illustrated in Figure~\ref{fig:compare}. Both Mamba and WebRTC use a monitor to collect network conditions and video information and finally make decisions for the video encoder to adapt to the network dynamics. However, WebRTC splits the ABR process into two separate sub-processes, i.e., target bitrate decision and bitrate control, and the bitrate control is also fulfilled by two separate modules, i.e., rate control in the encoder and resolution adaptation externally, resulting in the incoordination issues previously discussed. In contrast, Mamba uses a MARL-based model to automatically generate joint ABR algorithms for the end-to-end optimization from observations to the decision of encoding factor configurations, which avoids these incoordination issues and consequently improves the ABR performance. Prior studies have used single-agent reinforcement learning to facilitate the target bitrate decision process~\cite{ars, loki}. However, the large discrete action space of the MABR task, e.g., 168 (7$\times$4$\times6$) combinations of the atomic action in this work, is difficult for a single agent to handle~\cite{MING2023281}.

To this aim, we model the MABR problem as a multi-agent task that is composed of multiple autonomous agents that interact with the same environment and concurrently influence it. In detail, $qua$, $res$, and $fr$ agents are introduced to be in charge of adapting the QP, resolution, and frame rate, respectively. From the perspective of an individual agent, the effects of other agents' actions are integrated into the environment. Consequently, during the training, the environment is non-stationary for each agent as the policies of other agents are updated continuously. To overcome the non-stationary issue, we have meticulously designed the training process of Mamba, more details about which are provided in §\ref{ssec:train_methodology}. In the following, we describe how we model the MABR problem as a typical MARL task and choose the appropriate algorithm to solve it by explaining some essential concepts in the field of multi-agent tasks.

\textbf{Local or Global States.}
The first concern lies in the locality of the states that are used for training. Local states can be observed by the agent in the inference, while global states are only available during the training. In the design of Mamba, the states input to an agent is predominantly local, except for the encoding factors that will be decided by other agents. For example, the resolution and the frame rate applied to next-interval video frames, are two global states for the $qua$ agent, since they are being decided by the $res$ agent and the $fr$ agent respectively, and are not available. Further details regarding the states used by Mamba are introduced in §\ref{ssec:s_a_nn}.

\textbf{Homogeneous or Heterogeneous Agents.}
In multi-agent tasks, the homogeneity of learning agents refers to that multiple agents interact with an environment described by the same state space and take actions from the same action space. Homogeneous agents can leverage parameter-sharing techniques to accelerate the training process. However, in the design of Mamba, multiple agents are used to make decisions on different encoding factors, resulting in their disparate action spaces. Considering the discrepancy of decision tasks for each agent, the time span of state observations also differs, which leads to different state spaces. To address the challenge of training inefficiency resulting from these disparities, this paper proposes the use of curriculum learning, which will be detailed in §\ref{ssec:train_methodology}.

\textbf{Cooperative or Competitive Task.}
A multi-agent task can be cooperative, competitive, or hybrid mode. The classification is based on whether the optimization goals of each agent are fully identical, fully distinct, or partially identical. In the MABR task, the common goal of all agents is to optimize the overall QoE by 
maximizing the video quality in both spatial and temporal domains and minimizing the end-to-end video delay at the same time. This optimization goal is achieved by using a shared reward function, which is introduced in §\ref{ssec:train_algorithm}. Thus, MABR represents a cooperative task with identical optimization goals for all participating agents.

To sum up, the MABR is a cooperative task involving a group of heterogeneous agents that take actions based on both local and global states. In this paper, we use a state-of-the-art MARL framework, MAPPO~\cite{mappo}, to learn an ABR policy for Mamba. MAPPO has been demonstrated to perform well on cooperative tasks and heterogeneous multi-agent problems~\cite{papoudakis2021benchmarking}. To train Mamba's agents, we adopt a centralized training and decentralized execution (CTDE) approach, where the actors make decisions based on their own local states and a shared critic provides a centralized {\it value} based on global states. In the inference, we only use actor networks to represent Mamba's joint policy, and thus only local states are needed.

\section{Mamba Design}
\label{sec:design}

This section first introduces the detailed design of each agent, including the state space, action space, and neural network structure. These designs clarify the responsibilities of each agent and allow them to complete their tasks appropriately (§\ref{ssec:s_a_nn}). We then introduce our carefully designed training methodology to cope with the complex multi-agent explorations, enabling the agents to cooperate with each other to achieve a satisfying QoE (§\ref{ssec:train_methodology}). Finally, we provide a detailed description of the training algorithm we use (§\ref{ssec:train_algorithm}). 

\subsection{State, Action and Neural Network Design}
\label{ssec:s_a_nn}
For each of Mamba's agents $k\in\left \{ qua,res,fr \right \}$, they take actions ${{a}_{t}^{k}}$ based on their local observations including network conditions and video information about the environment as input state ${{s}_{t}^{k}}$, according to its policy ${{\pi }_{\theta }^{k}}({{s}_{t}^{k}},{{a}_{t}^{k}})$ which is represented by neural networks with parameters $\theta$. 

{\bf State.} 
To enable Mamba to coordinate the bandwidth estimation and bitrate control and jointly make decisions on QP, resolution, and frame rate, we include the network conditions, encoder parameters, and video content complexity in the input states.
In detail, the state used by Mamba at step $t$ is defined as $s_t=(\vec{u_t},\vec{v_t},\vec{c_t},\vec{y_t},\vec{f_t},\vec{h_t},\vec{d_t},\vec{p_t})$, and each of them is a vector of values observed in past $K$ intervals (e.g., $\vec{u_t}=\{u_{t-K+1},\ldots,u_t\}$).  
Here we omit the superscript $k$ of the state which indexes the agent because every agent uses the same state variables and only the range of statistical time differs.
$u_t$ and $v_t$ respectively represent the average Spatial perceptual Information (SI) and average Temporal perceptual Information (TI) of the raw frames in the same interval counted from step $t-1$ to step $t$. We include these metrics to measure content complexity which significantly impacts the rate-distortion~\cite{ou2010perceptual,ma2011modeling}. We calculate them following the standard reported in ITU-R BT.1788~\cite{ITU-R_BT.1788}.
$c_t$, $y_t$, and $f_t$ denote the average rate factor value, resolution, and average frame rate, respectively, which are applied to all frames in the interval for encoding. All of them are obtained from the encoder settings.
$h_t$, $d_t$, and $p_t$ are the playback frame rate, average RTT, and packet loss rate calculated during this interval, respectively.
In this work, the interval is set to 0.1s for $qua$ and $fr$ agents and 1s for $res$ agent. The same settings also apply to the action interval for these agents.
Specifically, the playback frame rate is defined as the number of actually played frames divided by the interval time. 
In this work, we set $k=6$ to capture dynamics of observations in the past 6 intervals, with the time span of the state for both $qua$ and $fr$ agents as 0.6s  (6$\times$0.1s) and that for $res$ agent as 6s (6$\times$1s). 

{\bf Action.} 
The goal of the $qua$ agent is to decide the QP for encoding the frames in the next interval. Considering better perceptual quality can be reached by using constant rate factor (CRF) mode than constant QP mode when encoding multiple frames, we use the CRF encoding mode and dynamically adjust the rate factor in this work. In order to avoid dramatic picture quality fluctuations, we build the action set of $qua$ agent with incremental adjustments of rate factor, i.e., $a_t^{qua}\in{\{+8,+4,+2,0,-1,-2,-4\}}$.
Considering the GCC in the default WebRTC framework adopts a strategy that is conservative to improve the video bitrate for congestion avoidance and is more aggressive to decrease the video bitrate when congestion occurs, we set this asymmetric rate factor list to keep this characteristic.
For the $fr$ agent, we empirically configure the 7-level absolute frame rates as its action set, i.e., $a_t^{fr}\in{\{60,50,40,30,20,10,0\}}$.
Both of the above two agents collect states and perform decisions every 0.1s. The intuition behind this is that the agents need to gather adequate network state information at each decision point while responding promptly to network congestion. For the $res$ agent, we pick up four resolution configurations widely used in mainstream RTVC scenarios, i.e., $a_t^{res}\in{\{1440p,1080p,720p,540p\}}$. The aspect ratio of the video is kept at 16:9.
The $res$ agent collects states and performs decisions every 1s, as every resolution switch requires a restart of the video encoder.

\textbf{Neural networks.} 
Mamba utilizes the multi-agent actor-critic framework to generate its control policy. In this framework, separate neural networks are employed for each agent's actor and critic, and a shared critic neural network is additionally introduced to produce a centralized {\it value} signal. To focus on ABR framework optimization, we design a lightweight neural network architecture inspired by prior works~\cite{loki, ars}, which applies to all actor and critic networks in the multi-agent actor-critic framework. The state inputs are first passed into a gated recurrent unit (GRU) layer with 64 units to capture temporal features, which are then fed into two cascaded fully connected networks with 64 and 32 neurons respectively. All of the hidden layers use the rectified linear unit (ReLU) as the activation function. Finally, the output layer employs the SoftMax activation function to obtain the final probability distribution of action options.

\vspace{-14pt}
\subsection{Training Methodology}
\label{ssec:train_methodology}

\begin{figure}[t]
    \centering
    \setlength{\abovecaptionskip}{0.15cm}
    \includegraphics[width=0.95\linewidth]{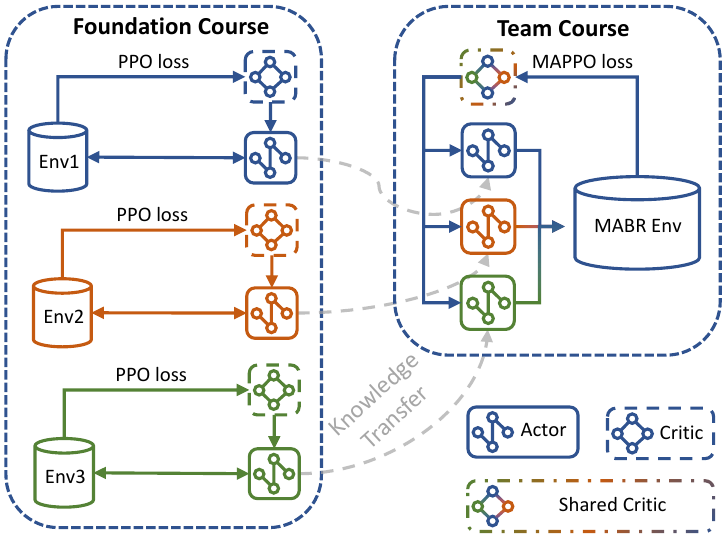}
    \caption{The two-stage curriculum learning that Mamba uses to train its ABR policy.}
    \label{fig:train}
    \vspace{-10pt}
\end{figure}

To support joint adaptation of the QP, resolution, and frame rate, Mamba uses MARL to learn an end-to-end multi-dimensional ABR policy. The most intuitive idea is to train Mamba's policy directly from scratch using the existing MARL algorithm. However, it may be very inefficient with unacceptable training time, due to the inherent non-stationarity problem of multi-agent tasks, as discussed in §\ref{sec:ma_design}. As a result, direct training from scratch can easily prevent an agent from exploring the optimal policy since it is unable to cooperate with other agents perfectly~\cite{hernandez2017survey, MING2023281}.
To address the issues, we introduce curriculum learning in the training of Mamba, which has been proven to be helpful in multi-agent training~\cite{from_few_to_more, cm3}. As shown in Figure~\ref{fig:train}, we design a two-stage curriculum for MABR learning: foundation course and team course.

{\bf Foundation course.} During the foundation course, a unique environment is established for each agent. For instance, for the $qua$ agent, the environment is transformed from non-stationary to stationary by fixing the resolution and the frame rate and blocking the policies of the other two agents. Consequently, the state inputs involving resolution and frame rate are masked, and the corresponding neurons are frozen in the neural networks during the training. Although the knowledge acquired by the agents in this environment may differ somewhat from the ultimate policy, their semantic experiences remain consistent~\cite{from_few_to_more}. In this stage, we train each agent's policy individually using the vanilla PPO algorithm~\cite{ppo}, and the training for each agent can be considered as a single-agent reinforcement learning process.

{\bf Team course.} The objective of the team course is to enable the agent to refine its own learned policy and gain the ability to collaborate with other agents. To achieve this, we initialize the Mamba using the weights obtained from the previous stage, which enables the Mamba to quickly acquire an amount of knowledge about the environment and prevents it from getting stuck in the early stage of training. In this stage, we use the complete state inputs and unfrozen neural networks for training. We utilize the MAPPO algorithm~\cite{mappo} to train Mamba's policy in this stage, and the basic training algorithm will be introduced in the following subsection.

\subsection{Basic Training Algorithm}
\label{ssec:train_algorithm}


{\bf Reward.} The goal of Mamba's RL agents is to maximize the expected cumulative discounted reward that they receive from the environment. Different from existing works that actually optimize the quality of service at the transport layer (e.g., throughput), we set the reward to directly reflect the user's QoE at the application layer, including the quality of the encoded video, the fluency of video playback, and end-to-end video delay. 
Specifically, the reward function could be formulated as follows:
\begin{equation}
\label{eq:reward}
    {{r}_{t}}=\lambda {q}_{t}+\nu {h}_{t}-\mu {d}_{t}.
\end{equation}
Here, ${{q}_{t}}$ is the perceptual video quality at step $t$. In order to reflect the picture quality more accurately and directly, we use the rate factor to calculate $q_t$ instead of video bitrate. 
${{h}_{t}}$ represents the playback frame rate, which directly affects the fluency of video playback.
${{d}_{t}}$ refers to the average frame delay (in seconds), which is used to approximate the end-to-end video delay between the sender and the receiver in a RTVC session. 
$\lambda$, $\nu$, and $\mu$ are weight coefficients that are used to adjust the bias of the policy. In this work, we set them to $1$, $8$, and $6$ empirically. Note that changing these three parameters did not affect the convergence of the model, but had a significant impact on the preference of the final policy.

\textbf{Loss function.} 
The loss function to train Mamba's policy is defined as:
\begin{equation}
    \begin{aligned}
    \mathcal{L} = & -\sum_{t} \sum_{k} \min \left(ratio_{\theta ,t}^{k}, \operatorname{clip}\left(ratio_{\theta, t}^{k}, 1-\epsilon, 1+\epsilon\right)\right)A_{t}^{GAE} \\ & +\beta\sum_{t} \sum_{k} H\left(\pi_{\theta}\left(s_{t}^{k}\right)\right),
    \end{aligned}
\end{equation}
where the first term is the widely-used clipped surrogate loss function, and the second term is the policy entropy which is added to avoid falling into sub-optimal policies at the start of the training. The advantage function $A_t^{GAE}$ is computed using the GAE~\cite{gae} method. This value measures the difference between the expected reward obtained when the agent picks an action $a_t$ in state $s_t$ deterministically, and the expected reward obtained through actions following the policy ${{\pi }_{\theta }}$ with the policy parameters $\theta$. In this formulation, $ratio_{\theta, t}^{k}$ and $H(\cdot)$ correspond to the surrogate objective and policy entropy, respectively. The index $k$ identifies a learning agent in Mamba, where $k\in\left \{ qua,res,fr \right \}$. The value of entropy weight parameter $\beta$ decreases when the reward no longer improves for 100 epochs. Further information regarding the technical details of the training algorithm is available in~\cite{mappo}.

\begin{figure}[t]
    \centering
    \setlength{\abovecaptionskip}{0.1cm}
    \includegraphics[width=\linewidth]{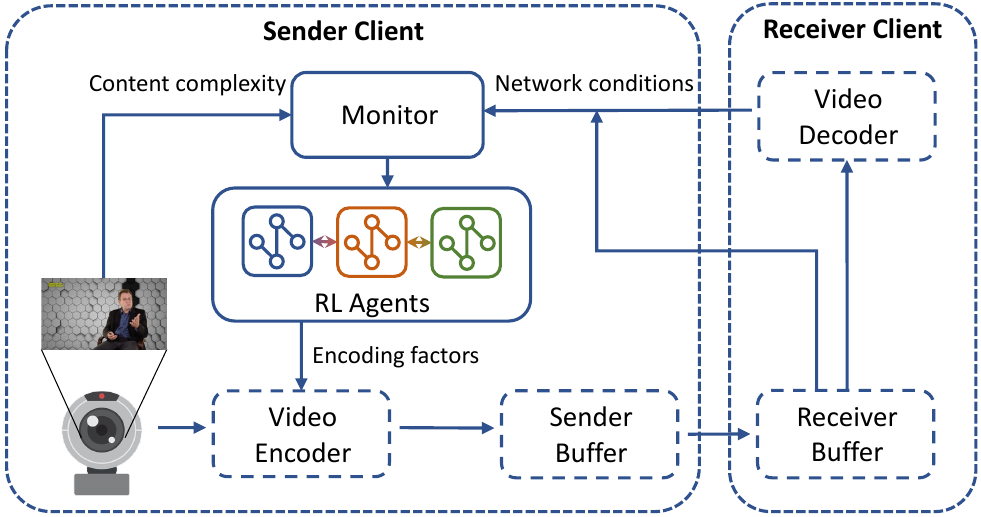}
    \caption{The system implementation of Mamba. We integrate an additional monitor module and the Mamba agents into the existing WebRTC system to build our testbed.}
    \label{fig:system}
    \vspace{-10pt}
\end{figure}

\section{Evaluation}
\label{sec:evaluation}
\subsection{Methodology}
\label{ssec:eva_methodology}
\textbf{Implementation.}
We implemented Mamba as an independent module to integrate into a WebRTC-compatible testbed, with the aim to replace the original congestion control and bitrate control modules at the system level. All of the experiments for observation (§\ref{ssec:observ}) and evaluation were conducted based on this testbed. Figure~\ref{fig:system} exhibits the seamless integration of Mamba into the existing WebRTC system. The modules marked by solid lines in the figure are added by Mamba, primarily consisting of Mamba agents and a monitor to gather state observations for Mamba. And the remaining modules indicated by dashed lines inherit from the original WebRTC system. The monitor module collects the observations of content complexity information and network conditions, which are subsequently input into Mamba's RL agents to make decisions on encoding factors for the video encoder. Additionally, we also implemented a state-of-the-art congestion control algorithm Loki~\cite{loki} in the testbed as a comparison to better evaluate Mamba.
In the implementation, we utilized CRF mode for Mamba and VBR mode for WebRTC and Loki (as they can only generate a target bitrate for the video encoder).
In the experiment, a sender client running on a PC connects to a receiver client running on another PC via a relay, in which the Linux traffic control tool is used to emulate different network conditions using network traces collected in the real world. At the sender client, a virtual camera is deployed to capture the prepared source videos, which are then processed to generate WebRTC-compatible video streams. This setup provides a controlled environment that enables us to compare the performances of different ABR approaches under the same network condition.

\textbf{Videos.}
In order to ascertain the impact of video content on ABR performance, a set of representative videos was chosen from the YouTube UGC video dataset~\cite{yt-ugc}, which comprises a diverse range of video genres such as games, video conferences, news interviews, and sports. These videos were concatenated into a continuous 2-minute sequence to create the test video source. This ensures the consistency of the video source for each algorithm during the evaluation process.

\begin{table}[t]
\setlength{\abovecaptionskip}{0.1cm}
  \centering
  \caption{Comparing Mamba with default WebRTC (GCC) and Loki on Belgium-4G, US-5G, Orca, and NYU-METS network trace datasets.}
  \resizebox{\columnwidth}{!}{%
    \begin{tabular}{cccccc}
    \toprule
    Dataset & Methods & \multicolumn{1}{p{4.78em}}{\begin{tabular}[c]{@{}c@{}}Video rate\\(kbps)\end{tabular}} & VMAF & \multicolumn{1}{p{3.39em}}{\begin{tabular}[c]{@{}c@{}}Delay\\(ms)\end{tabular}} & \multicolumn{1}{p{5.055em}}{\begin{tabular}[c]{@{}c@{}}Frame rate\\(fps)\end{tabular}} \\
    \toprule
    \multirow{3}[6]{*}{Orca~\cite{orca}} & WebRTC & 4053.02  & 48.60  & 124.70  & 55.56 \\
\cmidrule{2-6}      & Loki & 6800.42  & 50.15  & 159.48  & 48.30 \\
\cmidrule{2-6}      & Mamba & 4266.06  & 49.39  & 122.82  & 55.98 \\
    \midrule
    \multirow{3}[6]{*}{NYU-METS~\cite{nyu-mets}} & WebRTC & 4127.83  & 48.12  & 144.02  & 55.05 \\
\cmidrule{2-6}      & Loki & 7140.96  & 48.82  & 184.06  & 48.24 \\
\cmidrule{2-6}      & Mamba & 4966.93  & 53.91  & 137.85  & 54.79 \\
    \midrule
    \multirow{3}[6]{*}{Belgium-4G~\cite{Belgium}} & WebRTC & 4599.25 & 50.33 & 127.59 & 55.75 \\
\cmidrule{2-6}      & Loki & 9279.64 & 61.51 & 154.48 & 50.20 \\
\cmidrule{2-6}      & Mamba & 7023.10 & 60.90 & 125.34 & 55.86 \\
    \midrule
    \multirow{3}[6]{*}{US-5G~\cite{sigcomm'21}} & WebRTC & 9180.24 & 67.97 & 111.91 & 55.76 \\
\cmidrule{2-6}      & Loki & 18594.67  & 80.69  & 129.60  & 51.36 \\
\cmidrule{2-6}      & Mamba & 14294.29 & 80.44 & 115.08 & 55.98 \\
    \midrule
    \midrule
    \multirow{3}[6]{*}{\textbf{Overall}} & WebRTC & 5490.08  & 53.76  & 127.06  & 55.53 \\
\cmidrule{2-6}      & Loki & \textbf{10453.92}  & 60.29  & 159.91  & 49.53 \\
\cmidrule{2-6}      & Mamba & 7637.60  & \textbf{61.16}  & \textbf{125.27}  & \textbf{55.65} \\
    \bottomrule
    \end{tabular}%
    }
  \label{tab:in_lab_average}%
  \vspace{-8pt}
\end{table}%

\textbf{Network traces.} 
As with prior studies~\cite{loki,ars,concerto}, we conducted controlled laboratory simulations on several public network trace datasets, including:
(1) Orca~\cite{orca}: an AT\&T and T-Mobile cellular network trace dataset collected in New York City using mobile devices when the cellular user is walking, stationary, and riding a bus;
(2) NYU-METS~\cite{nyu-mets}: a set of LTE mobile network traces collected in New York City MTA on bus and subway riding;
(3) Belgium-4G~\cite{Belgium}: a 4G network trace dataset collected in Belgium on foot and in a variety of vehicles including bicycles, buses, trams, and trains;
(4) US-5G~\cite{sigcomm'21}: a set of 5G network traces collected in the US in both SA and NSA modes.
Obviously, the considered traces cover diverse access networks, regions, and scenarios, which best capture different network behaviors in the real world. Notably, each trace was post-processed with a granularity of 0.5s for bandwidth sampling. We randomly partitioned these traces into two sets, 80\% for training and the remaining 20\% for testing.

\textbf{Evaluation metrics.} In the field of RTVC, the user's QoE serves as the most basic metric to evaluate the quality of service. Since the QoE is susceptible to multiple factors from both transport and application layers, it is challenging to develop a definitive QoE model for assessment. To mitigate this issue, we use several popular QoE metrics to evaluate Mamba’s performance from various perspectives, including average video bitrate, average picture quality, average playback frame rate, and average delay. For the picture quality assessment, we use the Video Multi-method Assessment Fusion (VMAF~\cite{VMAF}), a widely recognized objective perceptual quality metric, to ensure fair comparisons. Previous studies~\cite{comyco} have reported that VMAF is more closely related to human perception than other metrics such as Peak Signal-to-Noise Ratio (PSNR), Structural Similarity (SSIM~\cite{SSIM}), etc. The video bitrate and picture quality are both captured at the sender client, which are used to represent the QoE sensation without considering the network-induced distortions. We use the average round-trip time (RTT) as an indicator to measure the interaction delay experienced throughout the entire session. Ultimately, an optimal ABR policy should prioritize the improvement of picture quality and playback frame rate while minimizing the delay to achieve a satisfactory QoE.

\begin{figure*}[t]
    \centering
    \setlength{\abovecaptionskip}{0.1cm}
    \includegraphics[width=\linewidth]{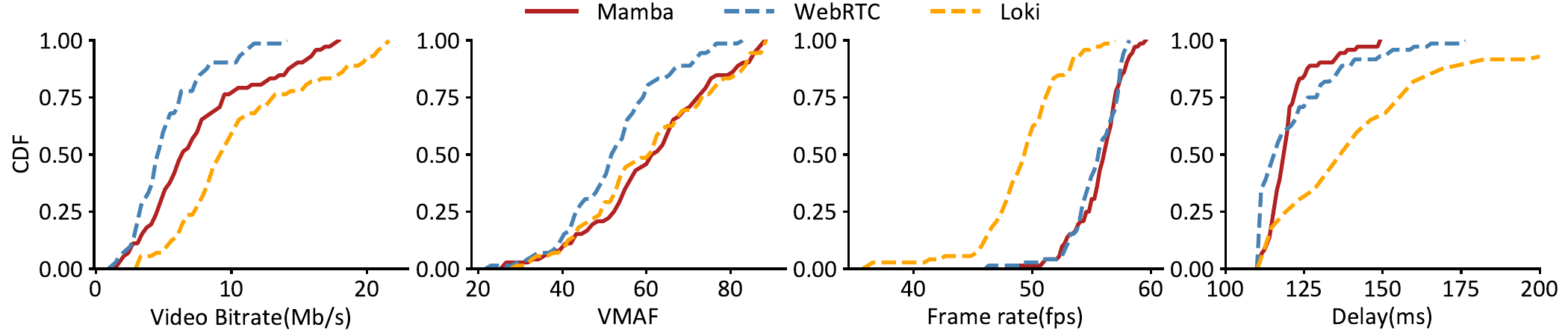}
    \caption{Comparing Mamba with the state-of-the-art WebRTC and Loki in the form of CDF distribution. The metrics of video bitrate, picture quality (in VMAF), playback frame rate, and delay are used to evaluate their performances. Results were collected on all network datasets.}
    \label{fig:in_lab_cdf}
    \vspace{-4pt}
\end{figure*}

\subsection{In-Lab Performance}
We evaluated the performance of Mamba and compared it with WebRTC and Loki using the methodology described in §\ref{ssec:eva_methodology}. The average results using all algorithms on different datasets are listed in Table~\ref{tab:in_lab_average}. Figure~\ref{fig:in_lab_cdf} provides more detailed results in the form of cumulative distribution functions (CDF). Based on the analysis of these results, we can yield three key findings.

Firstly, Mamba achieves the best overall performance on almost each QoE metric across different networks. Specifically, Mamba outperforms WebRTC and Loki on the metric of VMAF (by 13.7\% and 1.4\%), delay (by 1.4\% and 21.7\%), and playback frame rate (by 0.2\% and 12.4\%) with an acceptable video bitrate. In contrast, WebRTC tends to be conservative which is evidenced by its lowest video bitrate, leading to the worst picture quality in VMAF. And the aggressive Loki achieves the highest video bitrate (but not picture quality) at the cost of the largest delay and the lowest playback frame rate, which also fails to provide a satisfactory QoE for users. Furthermore, Mamba achieves consistently better performance under varying network conditions due to its advanced end-to-end multi-dimensional ABR policy. Figure~\ref{fig:in_lab_cdf} illustrates the absolute advantage of Mamba over other algorithms in the form of CDF.

Secondly, Mamba gains the superior capability of bandwidth adaptation by bridging the incoordination between transport and application layers. This is evidenced by significantly higher video bitrate (which results in a substantial improvement in picture quality) for Mamba compared to WebRTC, while still ensuring slightly better frame rates and delay. This demonstrates Mamba's capability to swiftly recover from network congestion while utilizing as much bandwidth as possible.

Thirdly, Mamba achieves a higher video quality in both spatial and temporal domains even at a relatively lower bitrate. As shown in Table~\ref{tab:in_lab_average}, Mamba achieves a higher VMAF and a higher playback frame rate compared to Loki using a lower video bitrate (roughly 3Mbps). The reason is that Mamba learns a joint adaptation policy of QP, resolution, and frame rate, which can dynamically choose an appropriate combination of these encoding factors to achieve desirable video quality.

We would like to emphasize that Mamba can change its policy tendency by adjusting the weights in the reward function (Equation~\eqref{eq:reward}) to suit diverse user preferences or different application platforms. For example, we can train another policy for Mamba using bigger $\nu$ and $\mu$ and smaller $\lambda$, leading to lower delay and higher playback frame rate (i.e., temporal quality) at the cost of lower picture quality (i.e., spatial quality).

\begin{figure}[t]
    \centering
    \setlength{\abovecaptionskip}{0.1cm}
    \subfigure[Stationary scene with WIFI network.]
    {
        \label{sfig:real_wifi}
        \begin{minipage}[ht]{.48\linewidth}
        \centering
            \includegraphics[width=\linewidth]{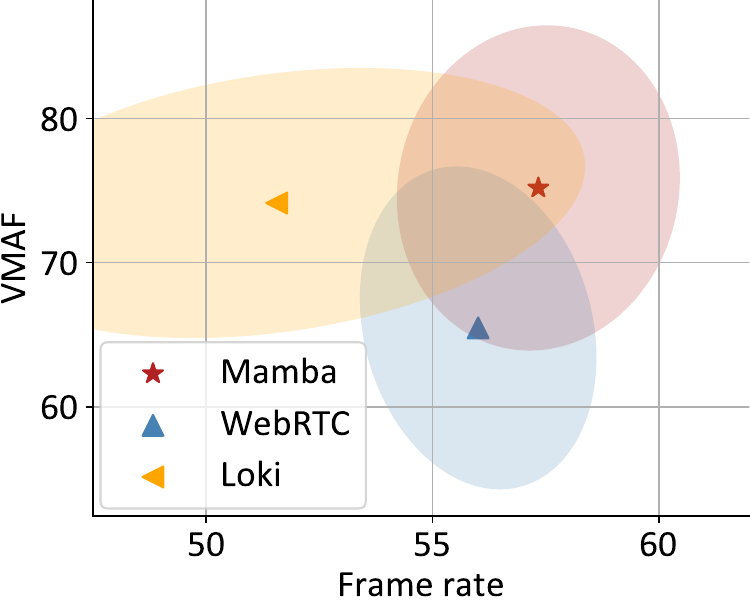}
        \end{minipage}
    }\hspace{-6pt}
    \subfigure[Mobility scene with 4G network.]
    {
        \label{sfig:real_4g}
        \begin{minipage}[ht]{.48\linewidth}
        \centering
            \includegraphics[width=\linewidth]{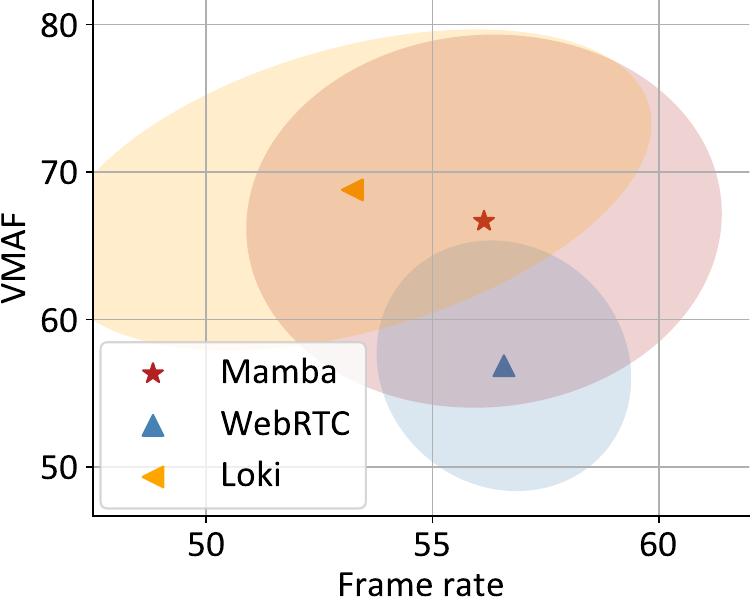}
        \end{minipage}
    }
    \caption{Real-world evaluation results of Mamba, WebRTC, and Loki in both stationary and moving scenarios.}
    \label{fig:real_world_results}
    \vspace{-6pt}
\end{figure}

\subsection{Real-World Performance}
To evaluate the generalization capability of Mamba, we conducted a series of real-world experiments covering a wide range of scenarios (stationary and moving) and network conditions (4G and WIFI). The stationary scenarios are comprised of accessing the campus WIFI network under a mixed mode of 2.4GHz and 5GHz within university buildings. Meanwhile, the moving scenarios involve accessing the 4G network while traveling by e-bike at a speed of approximately 20km/h or bus route in a city. All other experimental settings remained the same as described in the preceding section.

The results of the experiments are presented in Figure~\ref{fig:real_world_results}, which demonstrates the superior performance of Mamba compared to WebRTC and Loki. Mamba displayed better performance than WebRTC, with a more timely response to network congestion and efficient bandwidth utilization. This leads to significant improvements in picture quality (about 16.9\% improvement in VMAF) while maintaining high frame rates. Loki achieves a similar picture quality as Mamba while suffering significant frame rate degradation (about 4fps), which indicates more stalling. These results confirm the effectiveness of Mamba in RTVC applications.

\section{Discussion}
Although we implemented and evaluated the proposed Mamba specifically on the WebRTC platform to avoid complicated engineering issues, our method can also be applied to other RTVC applications which require high picture quality and low interaction delay, such as cloud gaming and virtual reality applications. Therefore, our contribution is not limited to optimizing the WebRTC and video conferencing experience. Furthermore, the MARL training algorithm used in this study is based on off-the-shelf advanced algorithms (e.g., MAPPO~\cite{mappo}) with state-of-the-art performance. Even when more advanced algorithms emerge in the field of MARL, our ABR approach will still be applicable.

\section{Related Work}
\label{sec:related_work}
\textbf{Real-time video communication.}
WebRTC is a widely used framework that utilizes rules-based GCC for congestion control and separate modules for rate control. However, the incoordination between the transport and application layers seriously degrades the performance. To address this issue, Salsify~\cite{salsify} was proposed to dynamically adjust the frame rate based on bandwidth estimation by accessing the internal state of the encoder. Unfortunately, Salsify's simple hand-crafted strategies and lack of support for resolution adaptation lead to poor performance. Furthermore, Salsify requires a specific encoder, making it difficult to deploy on existing RTVC platforms. Recently, several learning-based methods have been proposed to enhance congestion control performance. T-Gaming~\cite{ars} used RL to improve congestion control performance in cloud gaming, while Concerto~\cite{concerto} utilized imitation learning (IL) to improve congestion control performance in mobile video telephony. Furthermore, Loki~\cite{loki} proposed a hybrid method that combines RL and IL to achieve state-of-the-art congestion control performance. However, these learning-based methods still follow the same paradigm as WebRTC and suffer from the same limitations as previously mentioned.

\textbf{Multi-agent reinforcement learning.} Multi-agent reinforcement learning has been proven to have significant advantages in solving complex problems. Based on the training paradigm, it can be mainly classified into two categories: fully decentralized, and centralized training and decentralized execution (CTDE). IPPO~\cite{ippo} is a representative algorithm of the fully decentralized training methods, which is typically applied in scenarios where each agent can access global observation states and the agents are homogeneous (which can share parameters between each other). MAPPO~\cite{mappo} and MADDPG~\cite{maddpg} are representative algorithms of CTDE methods, where each agent usually can only observe local state values and use a shared reward signal for parameter updates. To better solve the problem of non-stationary environments in MARL, CTDE typically uses global state values as input to the critic for better training. The latter's characteristics are more suitable for the MABR task. In addition, to make training more efficient, curriculum learning was proposed in CM3~\cite{cm3} and DyMA-CL~\cite{from_few_to_more} to guide agents by incrementally increasing the difficulty of the agents' tasks, which has also been introduced in this work to improve the training efficiency.

\section{Conclusion}
We introduced and evaluated Mamba, a novel end-to-end multi-dimensional ABR framework for real-time video communication, which leverages multi-agent reinforcement learning to effectively coordinate the target bitrate decision and bitrate control processes and jointly determines the encoding factors including quantization parameters, frame rate, and resolution, resulting in an end-to-end bitrate adaptation that prioritizes QoE considerations. Through both in-lab and real-world tests in diverse scenarios, we demonstrate that Mamba outperforms existing state-of-the-art approaches across a wide range of networks and scenes.

\begin{acks}
This work was supported by the National Natural Science Foundation of China (62101241, 62231002) and the Jiangsu Provincial Double-Innovation Doctor Program (JSSCBS20210001). In addition, Yueheng Li especially wishes to thank Kobe Bryant, whose relentless competitive spirit and "Mamba Mentality" have been a source of great inspiration for him since childhood.
\end{acks}


\bibliographystyle{ACM-Reference-Format}
\balance
\bibliography{sample-base}


\begin{thebibliography}{34}


\ifx \showCODEN    \undefined \def \showCODEN     #1{\unskip}     \fi
\ifx \showDOI      \undefined \def \showDOI       #1{#1}\fi
\ifx \showISBNx    \undefined \def \showISBNx     #1{\unskip}     \fi
\ifx \showISBNxiii \undefined \def \showISBNxiii  #1{\unskip}     \fi
\ifx \showISSN     \undefined \def \showISSN      #1{\unskip}     \fi
\ifx \showLCCN     \undefined \def \showLCCN      #1{\unskip}     \fi
\ifx \shownote     \undefined \def \shownote      #1{#1}          \fi
\ifx \showarticletitle \undefined \def \showarticletitle #1{#1}   \fi
\ifx \showURL      \undefined \def \showURL       {\relax}        \fi
\providecommand\bibfield[2]{#2}
\providecommand\bibinfo[2]{#2}
\providecommand\natexlab[1]{#1}
\providecommand\showeprint[2][]{arXiv:#2}

\bibitem[Abbasloo et~al\mbox{.}(2020)]%
        {orca}
\bibfield{author}{\bibinfo{person}{Soheil Abbasloo}, \bibinfo{person}{Chen-Yu
  Yen}, {and} \bibinfo{person}{H.~Jonathan Chao}.}
  \bibinfo{year}{2020}\natexlab{}.
\newblock \showarticletitle{Classic Meets Modern: A Pragmatic Learning-Based
  Congestion Control for the Internet}. In
  \bibinfo{booktitle}{\emph{Proceedings of the Annual Conference of the ACM
  Special Interest Group on Data Communication on the Applications,
  Technologies, Architectures, and Protocols for Computer Communication}}
  (Virtual Event, USA) \emph{(\bibinfo{series}{SIGCOMM '20})}.
  \bibinfo{publisher}{Association for Computing Machinery},
  \bibinfo{address}{New York, NY, USA}, \bibinfo{pages}{632–647}.
\newblock
\showISBNx{9781450379557}
\urldef\tempurl%
\url{https://doi.org/10.1145/3387514.3405892}
\showDOI{\tempurl}


\bibitem[Chen et~al\mbox{.}(2019)]%
        {ars}
\bibfield{author}{\bibinfo{person}{Hao Chen}, \bibinfo{person}{Xu Zhang},
  \bibinfo{person}{Yiling Xu}, \bibinfo{person}{Ju Ren},
  \bibinfo{person}{Jingtao Fan}, \bibinfo{person}{Zhan Ma}, {and}
  \bibinfo{person}{Wenjun Zhang}.} \bibinfo{year}{2019}\natexlab{}.
\newblock \showarticletitle{T-Gaming: A Cost-Efficient Cloud Gaming System at
  Scale}.
\newblock \bibinfo{journal}{\emph{IEEE Transactions on Parallel and Distributed
  Systems}} \bibinfo{volume}{30}, \bibinfo{number}{12} (\bibinfo{year}{2019}),
  \bibinfo{pages}{2849--2865}.
\newblock
\urldef\tempurl%
\url{https://doi.org/10.1109/TPDS.2019.2922205}
\showDOI{\tempurl}


\bibitem[Cisco(2022)]%
        {video_traffic2022}
\bibfield{author}{\bibinfo{person}{Cisco}.} \bibinfo{year}{2022}\natexlab{}.
\newblock \bibinfo{booktitle}{\emph{Global - 2022 Forecast Highlights}}.
\newblock
\urldef\tempurl%
\url{https://www.cisco.com/c/dam/m/en_us/solutions/service-provider/vni-forecast-highlights/pdf/Global_2022_Forecast_Highlights.pdf}
\showURL{%
Retrieved March 1, 2022 from \tempurl}


\bibitem[de~Witt et~al\mbox{.}(2020)]%
        {ippo}
\bibfield{author}{\bibinfo{person}{Christian~Schroeder de Witt},
  \bibinfo{person}{Tarun Gupta}, \bibinfo{person}{Denys Makoviichuk},
  \bibinfo{person}{Viktor Makoviychuk}, \bibinfo{person}{Philip H.~S. Torr},
  \bibinfo{person}{Mingfei Sun}, {and} \bibinfo{person}{Shimon Whiteson}.}
  \bibinfo{year}{2020}\natexlab{}.
\newblock \bibinfo{title}{Is Independent Learning All You Need in the StarCraft
  Multi-Agent Challenge?}
\newblock
\newblock
\showeprint[arxiv]{2011.09533}~[cs.AI]


\bibitem[Fouladi et~al\mbox{.}(2018)]%
        {salsify}
\bibfield{author}{\bibinfo{person}{Sadjad Fouladi}, \bibinfo{person}{John
  Emmons}, \bibinfo{person}{Emre Orbay}, \bibinfo{person}{Catherine Wu},
  \bibinfo{person}{Riad~S. Wahby}, {and} \bibinfo{person}{Keith Winstein}.}
  \bibinfo{year}{2018}\natexlab{}.
\newblock \showarticletitle{Salsify: {Low-Latency} Network Video through
  Tighter Integration between a Video Codec and a Transport Protocol}. In
  \bibinfo{booktitle}{\emph{15th USENIX Symposium on Networked Systems Design
  and Implementation (NSDI 18)}}. \bibinfo{publisher}{USENIX Association},
  \bibinfo{address}{Renton, WA}, \bibinfo{pages}{267--282}.
\newblock
\showISBNx{978-1-939133-01-4}
\urldef\tempurl%
\url{https://www.usenix.org/conference/nsdi18/presentation/fouladi}
\showURL{%
\tempurl}


\bibitem[Gaetano et~al\mbox{.}(2017)]%
        {gcc}
\bibfield{author}{\bibinfo{person}{Gaetano}, \bibinfo{person}{Luca De~Cicco},
  \bibinfo{person}{Stefan Holmer}, {and} \bibinfo{person}{Saverio Mascolo}.}
  \bibinfo{year}{2017}\natexlab{}.
\newblock \showarticletitle{Congestion Control for Web Real-Time
  Communication}.
\newblock \bibinfo{journal}{\emph{IEEE/ACM Trans. Netw.}} \bibinfo{volume}{25},
  \bibinfo{number}{5} (\bibinfo{date}{oct} \bibinfo{year}{2017}),
  \bibinfo{pages}{2629–2642}.
\newblock
\showISSN{1063-6692}
\urldef\tempurl%
\url{https://doi.org/10.1109/TNET.2017.2703615}
\showDOI{\tempurl}


\bibitem[Hernandez-Leal et~al\mbox{.}(2017)]%
        {hernandez2017survey}
\bibfield{author}{\bibinfo{person}{Pablo Hernandez-Leal},
  \bibinfo{person}{Michael Kaisers}, \bibinfo{person}{Tim Baarslag}, {and}
  \bibinfo{person}{Enrique~Munoz De~Cote}.} \bibinfo{year}{2017}\natexlab{}.
\newblock \showarticletitle{A survey of learning in multiagent environments:
  Dealing with non-stationarity}.
\newblock \bibinfo{journal}{\emph{arXiv preprint arXiv:1707.09183}}
  (\bibinfo{year}{2017}).
\newblock


\bibitem[Huang et~al\mbox{.}(2019)]%
        {comyco}
\bibfield{author}{\bibinfo{person}{Tianchi Huang}, \bibinfo{person}{Chao Zhou},
  \bibinfo{person}{Rui-Xiao Zhang}, \bibinfo{person}{Chenglei Wu},
  \bibinfo{person}{Xin Yao}, {and} \bibinfo{person}{Lifeng Sun}.}
  \bibinfo{year}{2019}\natexlab{}.
\newblock \showarticletitle{Comyco: Quality-Aware Adaptive Video Streaming via
  Imitation Learning}. In \bibinfo{booktitle}{\emph{Proceedings of the 27th ACM
  International Conference on Multimedia}} (Nice, France)
  \emph{(\bibinfo{series}{MM '19})}. \bibinfo{publisher}{Association for
  Computing Machinery}, \bibinfo{address}{New York, NY, USA},
  \bibinfo{pages}{429–437}.
\newblock
\showISBNx{9781450368896}
\urldef\tempurl%
\url{https://doi.org/10.1145/3343031.3351014}
\showDOI{\tempurl}


\bibitem[ITU-R BT.1788(2007)]%
        {ITU-R_BT.1788}
ITU-R BT.1788 \bibinfo{year}{2007}\natexlab{}.
\newblock \bibinfo{booktitle}{\emph{Methodology for the subjective assessment
  of video quality in multimedia applications}}.
\newblock \bibinfo{type}{Standard}. \bibinfo{institution}{International
  Telecommunication Union}, \bibinfo{address}{Geneva, CH}.
\newblock


\bibitem[Jansen et~al\mbox{.}(2018)]%
        {10.1145/3199524.3199534}
\bibfield{author}{\bibinfo{person}{Bart Jansen}, \bibinfo{person}{Timothy
  Goodwin}, \bibinfo{person}{Varun Gupta}, \bibinfo{person}{Fernando Kuipers},
  {and} \bibinfo{person}{Gil Zussman}.} \bibinfo{year}{2018}\natexlab{}.
\newblock \showarticletitle{Performance Evaluation of WebRTC-Based Video
  Conferencing}.
\newblock  \bibinfo{volume}{45}, \bibinfo{number}{3} (\bibinfo{date}{mar}
  \bibinfo{year}{2018}), \bibinfo{pages}{56–68}.
\newblock
\showISSN{0163-5999}
\urldef\tempurl%
\url{https://doi.org/10.1145/3199524.3199534}
\showDOI{\tempurl}


\bibitem[Li et~al\mbox{.}(2023)]%
        {palette}
\bibfield{author}{\bibinfo{person}{Yueheng Li}, \bibinfo{person}{Hao Chen},
  \bibinfo{person}{Bowei Xu}, \bibinfo{person}{Zicheng Zhang}, {and}
  \bibinfo{person}{Zhan Ma}.} \bibinfo{year}{2023}\natexlab{}.
\newblock \bibinfo{title}{Improving Adaptive Real-Time Video Communication Via
  Cross-layer Optimization}.
\newblock
\newblock
\showeprint[arxiv]{2304.03505}


\bibitem[Lowe et~al\mbox{.}(2017)]%
        {maddpg}
\bibfield{author}{\bibinfo{person}{Ryan Lowe}, \bibinfo{person}{YI WU},
  \bibinfo{person}{Aviv Tamar}, \bibinfo{person}{Jean Harb},
  \bibinfo{person}{OpenAI Pieter~Abbeel}, {and} \bibinfo{person}{Igor
  Mordatch}.} \bibinfo{year}{2017}\natexlab{}.
\newblock \showarticletitle{Multi-Agent Actor-Critic for Mixed
  Cooperative-Competitive Environments}. In \bibinfo{booktitle}{\emph{Advances
  in Neural Information Processing Systems}},
  \bibfield{editor}{\bibinfo{person}{I.~Guyon}, \bibinfo{person}{U.~Von
  Luxburg}, \bibinfo{person}{S.~Bengio}, \bibinfo{person}{H.~Wallach},
  \bibinfo{person}{R.~Fergus}, \bibinfo{person}{S.~Vishwanathan}, {and}
  \bibinfo{person}{R.~Garnett}} (Eds.), Vol.~\bibinfo{volume}{30}.
  \bibinfo{publisher}{Curran Associates, Inc.}
\newblock
\urldef\tempurl%
\url{https://proceedings.neurips.cc/paper_files/paper/2017/file/68a9750337a418a86fe06c1991a1d64c-Paper.pdf}
\showURL{%
\tempurl}


\bibitem[Ma et~al\mbox{.}(2011)]%
        {ma2011modeling}
\bibfield{author}{\bibinfo{person}{Zhan Ma}, \bibinfo{person}{Meng Xu},
  \bibinfo{person}{Yen-Fu Ou}, {and} \bibinfo{person}{Yao Wang}.}
  \bibinfo{year}{2011}\natexlab{}.
\newblock \showarticletitle{Modeling of rate and perceptual quality of
  compressed video as functions of frame rate and quantization stepsize and its
  applications}.
\newblock \bibinfo{journal}{\emph{IEEE Transactions on Circuits and Systems for
  Video Technology}} \bibinfo{volume}{22}, \bibinfo{number}{5}
  (\bibinfo{year}{2011}), \bibinfo{pages}{671--682}.
\newblock


\bibitem[Mao et~al\mbox{.}(2017)]%
        {Pensieve}
\bibfield{author}{\bibinfo{person}{Hongzi Mao}, \bibinfo{person}{Ravi
  Netravali}, {and} \bibinfo{person}{Mohammad Alizadeh}.}
  \bibinfo{year}{2017}\natexlab{}.
\newblock \showarticletitle{Neural Adaptive Video Streaming with Pensieve}. In
  \bibinfo{booktitle}{\emph{Proceedings of the Conference of the ACM Special
  Interest Group on Data Communication}} (Los Angeles, CA, USA)
  \emph{(\bibinfo{series}{SIGCOMM '17})}. \bibinfo{publisher}{Association for
  Computing Machinery}, \bibinfo{address}{New York, NY, USA},
  \bibinfo{pages}{197–210}.
\newblock
\showISBNx{9781450346535}
\urldef\tempurl%
\url{https://doi.org/10.1145/3098822.3098843}
\showDOI{\tempurl}


\bibitem[Mei et~al\mbox{.}(2020)]%
        {nyu-mets}
\bibfield{author}{\bibinfo{person}{Lifan Mei}, \bibinfo{person}{Runchen Hu},
  \bibinfo{person}{Houwei Cao}, \bibinfo{person}{Yong Liu},
  \bibinfo{person}{Zifan Han}, \bibinfo{person}{Feng Li}, {and}
  \bibinfo{person}{Jin Li}.} \bibinfo{year}{2020}\natexlab{}.
\newblock \showarticletitle{Realtime mobile bandwidth prediction using LSTM
  neural network and Bayesian fusion}.
\newblock \bibinfo{journal}{\emph{Computer Networks}}  \bibinfo{volume}{182}
  (\bibinfo{year}{2020}), \bibinfo{pages}{107515}.
\newblock
\showISSN{1389-1286}
\urldef\tempurl%
\url{https://doi.org/10.1016/j.comnet.2020.107515}
\showDOI{\tempurl}


\bibitem[Ming et~al\mbox{.}(2023)]%
        {MING2023281}
\bibfield{author}{\bibinfo{person}{Fangzhu Ming}, \bibinfo{person}{Feng Gao},
  \bibinfo{person}{Kun Liu}, {and} \bibinfo{person}{Chengmei Zhao}.}
  \bibinfo{year}{2023}\natexlab{}.
\newblock \showarticletitle{Cooperative modular reinforcement learning for
  large discrete action space problem}.
\newblock \bibinfo{journal}{\emph{Neural Networks}}  \bibinfo{volume}{161}
  (\bibinfo{year}{2023}), \bibinfo{pages}{281--296}.
\newblock
\showISSN{0893-6080}
\urldef\tempurl%
\url{https://doi.org/10.1016/j.neunet.2023.01.046}
\showDOI{\tempurl}


\bibitem[Narayanan et~al\mbox{.}(2021)]%
        {sigcomm'21}
\bibfield{author}{\bibinfo{person}{Arvind Narayanan}, \bibinfo{person}{Xumiao
  Zhang}, \bibinfo{person}{Ruiyang Zhu}, \bibinfo{person}{Ahmad Hassan},
  \bibinfo{person}{Shuowei Jin}, \bibinfo{person}{Xiao Zhu},
  \bibinfo{person}{Xiaoxuan Zhang}, \bibinfo{person}{Denis Rybkin},
  \bibinfo{person}{Zhengxuan Yang}, \bibinfo{person}{Zhuoqing~Morley Mao},
  \bibinfo{person}{Feng Qian}, {and} \bibinfo{person}{Zhi-Li Zhang}.}
  \bibinfo{year}{2021}\natexlab{}.
\newblock \showarticletitle{A Variegated Look at 5G in the Wild: Performance,
  Power, and QoE Implications}. In \bibinfo{booktitle}{\emph{Proceedings of the
  2021 ACM SIGCOMM 2021 Conference}} (Virtual Event, USA)
  \emph{(\bibinfo{series}{SIGCOMM '21})}. \bibinfo{publisher}{Association for
  Computing Machinery}, \bibinfo{address}{New York, NY, USA},
  \bibinfo{pages}{610–625}.
\newblock
\showISBNx{9781450383837}
\urldef\tempurl%
\url{https://doi.org/10.1145/3452296.3472923}
\showDOI{\tempurl}


\bibitem[Ou et~al\mbox{.}(2010)]%
        {ou2010perceptual}
\bibfield{author}{\bibinfo{person}{Yen-Fu Ou}, \bibinfo{person}{Zhan Ma},
  \bibinfo{person}{Tao Liu}, {and} \bibinfo{person}{Yao Wang}.}
  \bibinfo{year}{2010}\natexlab{}.
\newblock \showarticletitle{Perceptual quality assessment of video considering
  both frame rate and quantization artifacts}.
\newblock \bibinfo{journal}{\emph{IEEE Transactions on Circuits and Systems for
  Video Technology}} \bibinfo{volume}{21}, \bibinfo{number}{3}
  (\bibinfo{year}{2010}), \bibinfo{pages}{286--298}.
\newblock


\bibitem[Papoudakis et~al\mbox{.}(2021)]%
        {papoudakis2021benchmarking}
\bibfield{author}{\bibinfo{person}{Georgios Papoudakis},
  \bibinfo{person}{Filippos Christianos}, \bibinfo{person}{Lukas Schäfer},
  {and} \bibinfo{person}{Stefano~V. Albrecht}.}
  \bibinfo{year}{2021}\natexlab{}.
\newblock \bibinfo{title}{Benchmarking Multi-Agent Deep Reinforcement Learning
  Algorithms in Cooperative Tasks}.
\newblock
\newblock
\showeprint[arxiv]{2006.07869}~[cs.LG]


\bibitem[Rassool(2017)]%
        {VMAF}
\bibfield{author}{\bibinfo{person}{Reza Rassool}.}
  \bibinfo{year}{2017}\natexlab{}.
\newblock \showarticletitle{VMAF reproducibility: Validating a perceptual
  practical video quality metric}. In \bibinfo{booktitle}{\emph{2017 IEEE
  International Symposium on Broadband Multimedia Systems and Broadcasting
  (BMSB)}}. \bibinfo{pages}{1--2}.
\newblock
\urldef\tempurl%
\url{https://doi.org/10.1109/BMSB.2017.7986143}
\showDOI{\tempurl}


\bibitem[Research and Markets(2022)]%
        {video_market2022}
\bibfield{author}{\bibinfo{person}{Research} {and} \bibinfo{person}{Markets}.}
  \bibinfo{year}{2022}\natexlab{}.
\newblock \bibinfo{booktitle}{\emph{Global Video Conferencing Market}}.
\newblock
\urldef\tempurl%
\url{https://www.researchandmarkets.com/reports/5415505}
\showURL{%
Retrieved March 1, 2022 from \tempurl}


\bibitem[Schulman et~al\mbox{.}(2015)]%
        {gae}
\bibfield{author}{\bibinfo{person}{John Schulman}, \bibinfo{person}{Philipp
  Moritz}, \bibinfo{person}{Sergey Levine}, \bibinfo{person}{Michael Jordan},
  {and} \bibinfo{person}{Pieter Abbeel}.} \bibinfo{year}{2015}\natexlab{}.
\newblock \bibinfo{title}{High-Dimensional Continuous Control Using Generalized
  Advantage Estimation}.
\newblock
\newblock
\urldef\tempurl%
\url{https://doi.org/10.48550/ARXIV.1506.02438}
\showDOI{\tempurl}


\bibitem[Schulman et~al\mbox{.}(2017)]%
        {ppo}
\bibfield{author}{\bibinfo{person}{John Schulman}, \bibinfo{person}{Filip
  Wolski}, \bibinfo{person}{Prafulla Dhariwal}, \bibinfo{person}{Alec Radford},
  {and} \bibinfo{person}{Oleg Klimov}.} \bibinfo{year}{2017}\natexlab{}.
\newblock \bibinfo{title}{Proximal Policy Optimization Algorithms}.
\newblock
\newblock
\urldef\tempurl%
\url{https://doi.org/10.48550/ARXIV.1707.06347}
\showDOI{\tempurl}


\bibitem[van~der Hooft et~al\mbox{.}(2016)]%
        {Belgium}
\bibfield{author}{\bibinfo{person}{Jeroen van~der Hooft},
  \bibinfo{person}{Stefano Petrangeli}, \bibinfo{person}{Tim Wauters},
  \bibinfo{person}{Rafael Huysegems}, \bibinfo{person}{Patrice~Rondao Alface},
  \bibinfo{person}{Tom Bostoen}, {and} \bibinfo{person}{Filip De~Turck}.}
  \bibinfo{year}{2016}\natexlab{}.
\newblock \showarticletitle{HTTP/2-Based Adaptive Streaming of HEVC Video Over
  4G/LTE Networks}.
\newblock \bibinfo{journal}{\emph{IEEE Communications Letters}}
  \bibinfo{volume}{20}, \bibinfo{number}{11} (\bibinfo{year}{2016}),
  \bibinfo{pages}{2177--2180}.
\newblock
\urldef\tempurl%
\url{https://doi.org/10.1109/LCOMM.2016.2601087}
\showDOI{\tempurl}


\bibitem[Wang et~al\mbox{.}(2020)]%
        {from_few_to_more}
\bibfield{author}{\bibinfo{person}{Weixun Wang}, \bibinfo{person}{Tianpei
  Yang}, \bibinfo{person}{Yong Liu}, \bibinfo{person}{Jianye Hao},
  \bibinfo{person}{Xiaotian Hao}, \bibinfo{person}{Yujing Hu},
  \bibinfo{person}{Yingfeng Chen}, \bibinfo{person}{Changjie Fan}, {and}
  \bibinfo{person}{Yang Gao}.} \bibinfo{year}{2020}\natexlab{}.
\newblock \showarticletitle{From Few to More: Large-Scale Dynamic Multiagent
  Curriculum Learning}.
\newblock \bibinfo{journal}{\emph{Proceedings of the AAAI Conference on
  Artificial Intelligence}} \bibinfo{volume}{34}, \bibinfo{number}{05}
  (\bibinfo{date}{Apr.} \bibinfo{year}{2020}), \bibinfo{pages}{7293--7300}.
\newblock
\urldef\tempurl%
\url{https://doi.org/10.1609/aaai.v34i05.6221}
\showDOI{\tempurl}


\bibitem[Wang et~al\mbox{.}(2019)]%
        {yt-ugc}
\bibfield{author}{\bibinfo{person}{Yilin Wang}, \bibinfo{person}{Sasi Inguva},
  {and} \bibinfo{person}{Balu Adsumilli}.} \bibinfo{year}{2019}\natexlab{}.
\newblock \showarticletitle{YouTube UGC Dataset for Video Compression
  Research}. In \bibinfo{booktitle}{\emph{2019 IEEE 21st International Workshop
  on Multimedia Signal Processing (MMSP)}}. \bibinfo{pages}{1--5}.
\newblock
\urldef\tempurl%
\url{https://doi.org/10.1109/MMSP.2019.8901772}
\showDOI{\tempurl}


\bibitem[Wang et~al\mbox{.}(2004)]%
        {SSIM}
\bibfield{author}{\bibinfo{person}{Zhou Wang}, \bibinfo{person}{A.C. Bovik},
  \bibinfo{person}{H.R. Sheikh}, {and} \bibinfo{person}{E.P. Simoncelli}.}
  \bibinfo{year}{2004}\natexlab{}.
\newblock \showarticletitle{Image quality assessment: from error visibility to
  structural similarity}.
\newblock \bibinfo{journal}{\emph{IEEE Transactions on Image Processing}}
  \bibinfo{volume}{13}, \bibinfo{number}{4} (\bibinfo{year}{2004}),
  \bibinfo{pages}{600--612}.
\newblock
\urldef\tempurl%
\url{https://doi.org/10.1109/TIP.2003.819861}
\showDOI{\tempurl}


\bibitem[WebRTC(2021)]%
        {webrtc}
\bibfield{author}{\bibinfo{person}{WebRTC}.} \bibinfo{year}{2021}\natexlab{}.
\newblock \bibinfo{booktitle}{\emph{Real-time communication for the web}}.
\newblock
\urldef\tempurl%
\url{https://webrtc.org/}
\showURL{%
Retrieved March 1, 2022 from \tempurl}


\bibitem[Wiegand et~al\mbox{.}(2003)]%
        {h264}
\bibfield{author}{\bibinfo{person}{T. Wiegand}, \bibinfo{person}{G.J.
  Sullivan}, \bibinfo{person}{G. Bjontegaard}, {and} \bibinfo{person}{A.
  Luthra}.} \bibinfo{year}{2003}\natexlab{}.
\newblock \showarticletitle{Overview of the H.264/AVC video coding standard}.
\newblock \bibinfo{journal}{\emph{IEEE Transactions on Circuits and Systems for
  Video Technology}} \bibinfo{volume}{13}, \bibinfo{number}{7}
  (\bibinfo{year}{2003}), \bibinfo{pages}{560--576}.
\newblock
\urldef\tempurl%
\url{https://doi.org/10.1109/TCSVT.2003.815165}
\showDOI{\tempurl}


\bibitem[Yang et~al\mbox{.}(2018)]%
        {cm3}
\bibfield{author}{\bibinfo{person}{Jiachen Yang}, \bibinfo{person}{Alireza
  Nakhaei}, \bibinfo{person}{David Isele}, \bibinfo{person}{Kikuo Fujimura},
  {and} \bibinfo{person}{Hongyuan Zha}.} \bibinfo{year}{2018}\natexlab{}.
\newblock \showarticletitle{Cm3: Cooperative multi-goal multi-stage multi-agent
  reinforcement learning}.
\newblock \bibinfo{journal}{\emph{arXiv preprint arXiv:1809.05188}}
  (\bibinfo{year}{2018}).
\newblock


\bibitem[Yin et~al\mbox{.}(2015)]%
        {MPC}
\bibfield{author}{\bibinfo{person}{Xiaoqi Yin}, \bibinfo{person}{Abhishek
  Jindal}, \bibinfo{person}{Vyas Sekar}, {and} \bibinfo{person}{Bruno
  Sinopoli}.} \bibinfo{year}{2015}\natexlab{}.
\newblock \showarticletitle{A Control-Theoretic Approach for Dynamic Adaptive
  Video Streaming over HTTP}. In \bibinfo{booktitle}{\emph{Proceedings of the
  2015 ACM Conference on Special Interest Group on Data Communication}}
  (London, United Kingdom) \emph{(\bibinfo{series}{SIGCOMM '15})}.
  \bibinfo{publisher}{Association for Computing Machinery},
  \bibinfo{address}{New York, NY, USA}, \bibinfo{pages}{325–338}.
\newblock
\showISBNx{9781450335423}
\urldef\tempurl%
\url{https://doi.org/10.1145/2785956.2787486}
\showDOI{\tempurl}


\bibitem[Yu et~al\mbox{.}(2021)]%
        {mappo}
\bibfield{author}{\bibinfo{person}{Chao Yu}, \bibinfo{person}{Akash Velu},
  \bibinfo{person}{Eugene Vinitsky}, \bibinfo{person}{Jiaxuan Gao},
  \bibinfo{person}{Yu Wang}, \bibinfo{person}{Alexandre Bayen}, {and}
  \bibinfo{person}{Yi Wu}.} \bibinfo{year}{2021}\natexlab{}.
\newblock \bibinfo{title}{The Surprising Effectiveness of PPO in Cooperative,
  Multi-Agent Games}.
\newblock
\newblock
\urldef\tempurl%
\url{https://doi.org/10.48550/ARXIV.2103.01955}
\showDOI{\tempurl}


\bibitem[Zhang et~al\mbox{.}(2021)]%
        {loki}
\bibfield{author}{\bibinfo{person}{Huanhuan Zhang}, \bibinfo{person}{Anfu
  Zhou}, \bibinfo{person}{Yuhan Hu}, \bibinfo{person}{Chaoyue Li},
  \bibinfo{person}{Guangping Wang}, \bibinfo{person}{Xinyu Zhang},
  \bibinfo{person}{Huadong Ma}, \bibinfo{person}{Leilei Wu},
  \bibinfo{person}{Aiyun Chen}, {and} \bibinfo{person}{Changhui Wu}.}
  \bibinfo{year}{2021}\natexlab{}.
\newblock \bibinfo{booktitle}{\emph{Loki: Improving Long Tail Performance of
  Learning-Based Real-Time Video Adaptation by Fusing Rule-Based Models}}.
\newblock \bibinfo{publisher}{Association for Computing Machinery},
  \bibinfo{address}{New York, NY, USA}, \bibinfo{pages}{775–788}.
\newblock
\showISBNx{9781450383424}
\urldef\tempurl%
\url{https://doi.org/10.1145/3447993.3483259}
\showURL{%
\tempurl}


\bibitem[Zhou et~al\mbox{.}(2019)]%
        {concerto}
\bibfield{author}{\bibinfo{person}{Anfu Zhou}, \bibinfo{person}{Huanhuan
  Zhang}, \bibinfo{person}{Guangyuan Su}, \bibinfo{person}{Leilei Wu},
  \bibinfo{person}{Ruoxuan Ma}, \bibinfo{person}{Zhen Meng},
  \bibinfo{person}{Xinyu Zhang}, \bibinfo{person}{Xiufeng Xie},
  \bibinfo{person}{Huadong Ma}, {and} \bibinfo{person}{Xiaojiang Chen}.}
  \bibinfo{year}{2019}\natexlab{}.
\newblock \showarticletitle{Learning to Coordinate Video Codec with Transport
  Protocol for Mobile Video Telephony}. In \bibinfo{booktitle}{\emph{The 25th
  Annual International Conference on Mobile Computing and Networking}} (Los
  Cabos, Mexico) \emph{(\bibinfo{series}{MobiCom '19})}. Article
  \bibinfo{articleno}{29}, \bibinfo{numpages}{16}~pages.
\newblock
\urldef\tempurl%
\url{https://doi.org/10.1145/3300061.3345430}
\showDOI{\tempurl}


\end{thebibliography}

\appendix

\end{document}